\documentclass[aps,prev,twocolumn,preprintnumbers,floatfix,nofootinbib]{revtex4-1}

\usepackage[T1]{fontenc} 
\usepackage[utf8]{inputenc}
\usepackage[dvipdfm]{graphicx}
\usepackage{mathrsfs}
\usepackage{amssymb}
\usepackage{verbatim}
\usepackage{color}
\usepackage[dvipsnames]{xcolor}
\usepackage{multirow}
\usepackage{amsmath}
\usepackage{subfig}
 \usepackage{slashed} 
\usepackage[normalem]{ulem}
\usepackage{sidecap}
\usepackage{hyperref}
\definecolor{teal}{rgb}{0.1,0.4,0.1}

\hypersetup{
    colorlinks = true,
    citecolor = {teal},
    linkcolor = {teal},
    urlcolor = {teal},
}

\newcommand{\lslashslash}{%

  \raisebox{0.8ex}{%
    \scalebox{.7}{%
      \rotatebox[origin=c]{18}{$-$}%
    }%
  }%
}
\newcommand{\lslash}{%
  {%
   \vphantom{d}%
   \ooalign{\kern-.1em\smash{\lslashslash}\hidewidth\cr${\rm l}$\cr}%
   \kern.05em
  }%
}

\newcommand{\beq}{\begin{equation}}
\newcommand{\eeq}{\end{equation}}
\newcommand{\bea}{\begin{eqnarray}}
\newcommand{\eea}{\end{eqnarray}}

\def\to{\rightarrow}

\usepackage[parfill]{parskip}
\begin{document}

\preprint{FTPI-MINN-21-04}
\preprint{UMN-TH-4011/21}
\preprint{IFT-UAM/CSIC-21-32}

\vspace*{1mm}

\title{Slow and Safe Gravitinos} 

\author{Emilian Dudas$^{a}$}
\email{emilian.dudas@polytechnique.edu}
\author{Marcos~A.~G.~Garcia$^{b}$}
\email{marcosa.garcia@uam.es}
\author{Yann Mambrini$^{c}$}
\email{yann.mambrini@th.u-psud.fr}
\author{Keith A. Olive$^{d}$}
\email{olive@umn.edu}
\author{Marco Peloso$^{e,f}$}
\email{marco.peloso@pd.infn.it}
\author{Sarunas~Verner$^{d}$}
\email{nedzi002@umn.edu}

\affiliation{$^a$ Centre de Physique Th{\'e}orique, {\'E}cole Polytechnique, CNRS and IP Paris, 91128 Palaiseau Cedex, France}

\affiliation{$^b$ Instituto de F\'isica Te\'orica (IFT) UAM-CSIC, Campus de Cantoblanco, 28049, Madrid, Spain}

\affiliation{${}^d $ Universit\'e Paris-Saclay, CNRS/IN2P3, IJCLab, 91405 Orsay, France}
\affiliation{$^e$William I. Fine Theoretical Physics Institute, School of
 Physics and Astronomy, University of Minnesota, Minneapolis, MN 55455,
 USA}
\affiliation{${}^f$ Dipartimento di Fisica e Astronomia “Galileo Galilei” Universit\`a di Padova, 35131 Padova, Italy  }

\affiliation{${}^g$ INFN, Sezione di Padova, 35131 Padova, Italy  }

\date{\today}

\begin{abstract} 
It has been argued that supergravity models of inflation
with vanishing sound speeds, $c_s$, lead to an unbounded growth in the production rate of gravitinos. We consider several models of inflation to delineate the conditions for which $c_s = 0$. In models with unconstrained superfields, we argue that the mixing of the goldstino and inflatino in a time-varying background prevents the uncontrolled production of the longitudinal modes. This conclusion is unchanged if there is a nilpotent field associated with supersymmetry breaking with constraint ${\bf S^2} =0$, i.e. sgoldstino-less models. Models with a second orthogonal constraint, ${\bf S(\Phi-\bar{\Phi})} =0$, where $\bf{\Phi}$ is the inflaton superfield, which eliminates the inflatino, may suffer from the over-production of gravitinos. 
However, we point out that these models may be problematic if this constraint originates from a UV Lagrangian, as this may require using higher derivative operators. These models may also exhibit other pathologies such as $c_s > 1$, which are absent in theories with the single constraint or unconstrained fields. 

\end{abstract}

\vskip 1cm

\maketitle

\setcounter{equation}{0}

\section{I. Introduction}

Although supersymmetry has so far eluded discovery at the LHC \cite{ATLAS20,CMS20}, it remains an important component for building models
beyond the Standard Model, particularly when going to grand unified scales and beyond. It is well known that supersymmetry can provide for the naturalness of the weak scale \cite{Maiani:1979cx}, the unification of gauge couplings \cite{Ellis:1990zq}, and a cold dark matter candidate \cite{gold,ehnos}. 
As supergravity~\cite{cremmer,susy} is the extension of supersymmetry which includes gravity, it is the
natural framework for any cosmological scenario involving supersymmetry.
Thus, the gravitino also becomes a component in post-inflationary cosmology. 

The gravitino has long since been a cosmological headache \cite{prob,sw,eln,ego}.
It has the potential to overclose the universe if it is stable
(and hence a dark matter candidate), or lead to an overabundance of the lightest supersymmetric particle if it decays. While inflation may dilute any primordial gravitinos \cite{eln}, they are reproduced during reheating after inflation \cite{nos,ehnos,kl,ekn,oss,Juszkiewicz:gg,mmy,Kawasaki:1994af,Moroi:1995fs,enor,Giudice:1999am,bbb,cefo,kmy,stef,Pradler:2006qh,ps2,rs,kkmy,egnop,Garcia:2017tuj,Eberl:2020fml}. Decaying gravitinos may also upset the concordance of big bang nucleosynthesis \cite{foyy} and these constraints result in limits on the gravitino abundance which ultimately translate into limits on the reheat temperature \cite{bbn,cefo,ceflos175,Kawasaki:1994af,stef}.

In addition, there is also a non-thermal contribution to the gravitino abundance stemming from the non-perturbative decay of the inflaton when supersymmetry is broken during inflation \cite{Kallosh:1999jj,Giudice:1999yt,Giudice:1999am,Kallosh:2000ve,Nilles:2001ry,Nilles:2001fg,Ema:2016oxl}. However, it was shown that what is produced in this process is the goldstino, that is, the combination of fermions contributing to supersymmetry breaking at the time of their production.
At the inflationary scale, the goldstino is typically dominated by the inflatino, the fermionic partner of the inflaton. The longitudinal component of the gravitino at low energy is associated with the true goldstino which may be unrelated to the inflatino produced in reheating \cite{Nilles:2001ry,Nilles:2001fg}. 
The production of inflatinos may be problematic, particularly if the mass of the inflatino is less than the mass of the inflaton, but may be suppressed and the final gravitino abundance is model dependent \cite{Nilles:2001my}.

Recently, it has been claimed that if the sound speed, $c_s$, for gravitinos 
were to vanish, there would be a catastrophic non-thermal production of slow gravitinos \cite{Kolb:2021xfn,Kolb:2021nob}.
This is due to the lack of suppression in the production of high momentum modes in the helicity 1/2 spectrum. 
While $c_s$ may indeed vanish in certain models of inflation derived from supergravity, as in the case of the non-thermal production of the longitudinal mode, mixing with other fermionic fields whose scalar partners have non-vanishing time derivatives precludes the runaway production of gravitinos.
Nevertheless, in some realizations of non-linear supergravity with constrained fields, the inflatino may not be present in the spectrum, and indeed catastrophic production may occur.

In this paper, we consider the conditions which lead to $c_s = 0$. We look at several 
models of inflation in supergravity theories 
and distinguish those for which $c_s = 0$ at some point in the post-inflationary oscillation period of the inflaton. We stress that $c_s = 0$ is not a generic feature of supergravity inflation models. Further we argue that even if $c_s = 0$, in the absence of constrained fields, catastrophic gravitino production does {\em not} occur. Indeed even in constrained models with a nilpotent field (eliminating a scalar from the spectrum) \cite{Rocek:1978nb,Komargodski:2009rz,Antoniadis:2014oya,Ferrara:2014kva,Kallosh:2014via,DallAgata:2014qsj,Dudas:2015eha,Ferrara:2015gta,Ferrara:2015cwa}, there is no catastrophic production.  Only when a second, orthogonal constraint \cite{Ferrara:2015tyn,Carrasco:2015iij,DallAgata:2015zxp,Kallosh:2016hcm,DallAgata:2016syy} which eliminates, the pseudoscalar, fermionic and auxiliary partners of the inflaton, can catastrophic production occur when $c_s = 0$. In \cite{hase}, only models with both constraints were considered leading to a truncated equation of motion for the longitudinal component of the gravitino \cite{Ferrara:2015tyn} and explosive particle production.  While the nilpotent constraint can easily be viewed as the infrared limit of a heavy sgoldstino (the scalar field associated with supersymmetry breaking), the orthogonal constraint appears to require higher derivative operators in the action \cite{DallAgata:2016syy}. Indeed we would argue that the constraints from gravitino production signal a further problem for the orthogonal constraint rather than imposing conditions on general models of inflation in supergravity.

In what follows, we first consider in Section \ref{sec:theta} the calculation of the sound speed in supergravity theories. It is easy to see that in models with a single chiral field, $c_s = 1$. We next consider in Section \ref{sec:slow} several models of inflation with supersymmetry breaking showing that $c_s$ is quite model dependent.
In some cases $c_s = 0$ occurs at some specific points during the oscillation of the inflaton, while in other often studied models, $c_s \simeq 1$ always. In Section \ref{sec:theta-Y}, we reconsider the equations of motion for the longitudinal component of the gravitino. We show clearly that even when $c_s = 0$, there is no enhanced production of gravitinos. The nilpotency conditions are discussed in Section \ref{sec:nil}. With a single nilpotent field, our conclusions are left unchanged.  Only when a second orthogonal condition is imposed, and the inflatino is eliminated from the spectrum, is there the potential for explosive gravitino production.  A discussion and our conclusions are given in Section \ref{sec:conclusions}.

\section{Gravitino sound speed in supergravity}
\label{sec:theta}

It is argued in \cite{Kolb:2021xfn,Kolb:2021nob} that the vanishing of the gravitino sound speed is accompanied by a catastrophic gravitino overproduction during the first set of  inflaton oscillations. The mathematical reason for this statement can be understood from the linearized theory of an uncoupled longitudinal gravitino. The gravitino has a dispersion relation of the type $\omega_k^2=c_s^2 k^2 + a^2 m^2(t)$, $a$ being the scale factor and $k$ the comoving momentum. As is well known from studies of preheating \cite{preheating,stb, kt,Greene:1997fu}, a non-adiabatic change of the frequency, $\omega_k' / \omega_k^2 \gg 1$ (where prime denotes derivative with respect to conformal time), results in significant particle production. Normally, the $k^2$ term ensures that the variation is adiabatic in the deep UV, so that modes of high momenta remain in their vacuum state. This is not the case if $c_s^2 =0$. In fact, the production obtained in \cite{hase,Kolb:2021xfn,Kolb:2021nob} is formally divergent in the UV, and it is presumably regulated by higher-order operators. This would in any case result in the production of an unacceptably large gravitino abundance. 

As we shall see in Section \ref{sec:theta-Y}, this problem is typically not present in multi-field models, since in this case the longitudinal gravitino is coupled already at the linearized level to a linear combination $\Upsilon$ of the fermions in the theory \cite{Kallosh:2000ve,Nilles:2001fg}. The gravitino momentum term is then replaced by a matrix in field space, that is non-singular when $c_s^2 = 0$. Therefore, the physical eigenvalues of the system do not vanish, and the catastrophic production in the UV is avoided. 

We begin by first computing $c_s$ in supergravity. 
The equations for the transverse and longitudinal gravitino component presented in this work follow the formalism of Ref.~\cite{Kallosh:2000ve}. In that work, the FLRW line element is chosen according to $ds^2 = a^2 \left( \tau \right) \left[ - d \tau^2 + d \vec{x}^2 \right]$, where $\tau$ is the conformal time. We denote by $\gamma^\mu$ the gamma matrices in flat spacetime. We work in units of $M_p =1$, where $M_p$ is the reduced Planck mass. 

The equation of motion for the transverse gravitino component is 
\begin{eqnarray}
\left( \gamma^0 \partial_0 + i \gamma^i k_i + a m_{3/2} \right) \psi^T = 0 
\label{eq-psi32}
\end{eqnarray}
where $\vec{k}$ is the comoving momentum, and $\partial_0$ is the derivative with respect to the conformal time. The propagation of the 
transverse mode of the gravitino
is not directly affected by the supersymmetry breaking mechanism, and obeys a classical Dirac equation with a speed of sound $c_s=1$. The gravitino mass is given by 
\begin{equation}
m_{3/2} = {\rm e}^{K/2} \, \left\vert W \right\vert \;, 
\end{equation}
where $K$ and $W$ denote the K\"ahler potential and the superpotential, respectively. 

The equation for the longitudinal gravitino will be presented and discussed in Section \ref{sec:theta-Y}. Here we discuss one important difference with Eq.~(\ref{eq-psi32}) for the longitudinal component. Namely, the momentum-dependent term in the equation for the longitudinal component is multiplied by a function of the background scalar fields, that, once squared, is identified as the square of the longitudinal gravitino sound speed $c_s^2$,
\beq
c_s^2 = \frac{\left(p-3 m_{3/2}^{2}\right)^{2}}{\left(\rho+3 m_{3/2}^{2} \right)^{2}}+\frac{4 \dot{m}_{3/2}^2}{\left(\rho+3 m_{3/2}^{2} \right)^{2}} \, ,
\label{Eq:generic32}
\eeq
where $p$ is the pressure, $\rho$ is the energy density, and dot denotes the derivative with respect to cosmological time. 
Ref.~\cite{Kolb:2021xfn} provides a rather compact expression for this quantity in a supergravity model, namely 
\begin{equation} 
c_s^2 = 1 - \frac{4}{
\left(\vert\dot{\varphi}\vert^2 + \vert F \vert^2 \right)^2} \, 
\left\{ 
 |\dot{\varphi}|^2  |F|^2  - 
\left\vert \dot{\varphi}  \cdot F^*  \right\vert^2 \right\}  \;, 
\label{vs2}
\end{equation} 
where $\varphi$ is the multiplet of scalar fields in the model, and the $F$-term is  
\begin{equation}
F^i \equiv {\rm e}^{K/2} K^{ij^*} \, D_{j^*} W^* \;, 
\label{DWF}
\end{equation}
where, using standard supergravity notation, $K^{ij^*}$ is the inverse of the K\"ahler metric 
\begin{equation}
K_{ij^*} \equiv \frac{\partial^2 K}{\partial \varphi^i \, \partial \varphi^{j*}} \;, 
\label{Kij}
\end{equation} 
while 
\begin{equation}
D_i W \equiv \frac{\partial W}{\partial \varphi^i} +  \frac{\partial K}{\partial \varphi^i} \, W \,. 
\end{equation} 
The dot operator in Eq.~(\ref{vs2}) denotes a  scalar product with the K\"ahler metric (\ref{Kij}), namely $\vert \dot{\varphi} \vert^2 = \dot{\varphi}^i \, K_{ij^*} \, \dot{\varphi}^{j*}$, and analogously for the other terms. 
Notice that due to the Cauchy-Schwarz type inequality
$|\dot{\varphi}|^2  |F|^2  \geq 
\left\vert \dot{\varphi}  \cdot F^*  \right\vert^2$, causality $c_s \leq 1$ is always guaranteed to hold.

In the case of a single chiral superfield, the two terms in the curly bracket in Eq. (\ref{vs2}) are equal to each other, and therefore $c_s=1$.
Thus, $c_s^2 \simeq 1$ is expected whenever a single field dominates the kinetic energy and supersymmetry breaking in the model. 

Ref. \cite{Kolb:2021xfn} considered the case in which multiple fields are relevant, and they conspire to give a vanishing or nearly vanishing $c_s^2$. 
This can be achieved if 
\begin{equation}
\dot{\varphi}  \cdot F^* = 0 \;\;{\rm and}\;\; 
\dot{\varphi}  \cdot \dot{\varphi}^* = 
F \cdot F^* \;\; \Rightarrow \;\; 
c_s^2 = 0 \;. 
\label{vs0}
\end{equation} 
Note that the first of these conditions, $\dot{\varphi}  \cdot F^* = 0$, is realized whenever the gravitino mass is constant. 
These conditions can be satisfied during inflaton oscillations after inflation. Typically, the inflaton dominates the kinetic energy, so the condition $\dot{\varphi}  \cdot F^* \simeq 0 $ generically requires that the $F$-term associated with the inflaton is small. Barring cancellations, this would typically require that both $W$ and $\frac{\partial W}{\partial \phi}$, where $\phi$ denotes the inflaton field, are small. We note that the potential energy is given by 
\begin{equation}
V = F \cdot F^* - 3 \, {\rm e}^K \, \left\vert W \right\vert^2 \;, 
\end{equation}
which we can approximate by $V \simeq F \cdot F^*$ if $W$ is small. Then the second condition in (\ref{vs0}) simply demands that the kinetic and the potential energy are equal to each other,  which happens twice per period of the inflaton oscillations. In Section \ref{sec:slow},  we discuss several supergravity models of inflation where these conditions are and are not achieved.

\section{Gravitino sound speed in specific models}
\label{sec:slow}

In this section we consider several specific supergravity models where $c_s^2$ is very small, or is of order one, to emphasize what aspects of the model can lead to a slow gravitino. 

We start our discussion from a model constructed in \cite{Kolb:2021xfn}, where $\Phi$ is the inflaton superfield while $S$ is a superfield responsible for supersymmetry breaking. Ref.  \cite{Kolb:2021xfn} imposes that this field is nilpotent, $S^2=0$. To study the relevance of this assumption we instead use a strong stabilization mechanism for $S$ \cite{ekn3,strongpol,dlmmo,ADinf,ego,eno7,egno4,Ellis:2020xmk} (see also Section \ref{sec:nil} below): 
\begin{eqnarray}
K &=& - \frac{\left( \Phi - {\bar \Phi} \right)^2}{2} + \frac{S \, {\bar S}}{ 1 + \frac{m^2}{M^2} \left\vert \Phi \right\vert^2} 
- \frac{\left( S \, {\bar S} \right)^2}{\Lambda^2} \;,  \nonumber\\ 
W &=& M \, S + W_0 \;. 
\label{kolb-Lambda}
\end{eqnarray}

The resulting potential is extremized along the real directions $\Phi = {\bar \Phi} = \frac{\phi}{\sqrt{2}} \;,\; S = {\bar S} = \frac{s}{\sqrt{2}}$. The minimum of the potential with respect to $s$ is $\phi-$dependent and given by: 
\begin{equation}
\left\langle s \right\rangle_\phi = \frac{\Lambda^2}{\sqrt{6} \left( \frac{m^2 \phi^2}{2 M^2}+1\right)^2} + \mathcal{O} \left( \Lambda^4 \right) \;. 
\label{smin-Kolb}
\end{equation} 
We see that, as is typical for strong stabilization, $\langle s \rangle_\phi$ vanishes in the limit $\Lambda \to 0$. Inserting this into the potential, leads to %
\begin{equation}
V = \frac{m^2 \phi^2}{2} + \frac{M^2}{3} \left( 1 - \frac{1}{\left(\frac{m^2 \phi^2}{2 M^2}+1\right)^2} \right) \Lambda^2 + \mathcal{O} \left( \Lambda^4 \right) \;. 
\label{Vmin-Kolb}
\end{equation}
In both Eqs.~(\ref{smin-Kolb}) and (\ref{Vmin-Kolb}), the parameter $W_0$ has been set to $W_0 = \frac{M}{\sqrt{3}} \left( 1 - \frac{\Lambda^2}{6} + \mathcal{O} \left( \Lambda^4 \right) \right)$, so to have a vanishing potential in the minimum at $\phi =0$. In the minimum, the gravitino mass is given by
\begin{equation}
m_{3/2} \Big\vert_{\phi = 0} = \frac{M}{\sqrt{3}} + \mathcal{O} \left( \Lambda^2 \right) \;. 
\label{m32-min-Kolb}
\end{equation} 
From Eqs.~(\ref{Vmin-Kolb}) and (\ref{m32-min-Kolb}) we see that $m$ corresponds to the inflaton mass, while (assuming gravity mediation), $M$ to the supersymmetry breaking scale in the model. Therefore, the ratio $(m/M)^2 \sim (m/m_{3/2})^2$
appearing in the K\"ahler potential is typically much greater than 1 in models with weak-scale supersymmetry breaking. 

Evaluating the quantities entering in Eq.~(\ref{vs2}) leads to 
\begin{eqnarray}
\dot{\varphi}^i &\simeq& \frac{\dot{\phi}}{\sqrt{2}} \left\{ 1 ,\, 
-  \frac{8 \sqrt{2} \Lambda^2 M^4}{\sqrt{3} m^4 \phi^5}  \right\} \;,
\nonumber\\ 
F^i &\simeq& \left\{ \sqrt{\frac{2}{3}} \, \frac{2 M^5 \Lambda^2}{m^4 \phi^5} ,\, \frac{m^2 \phi^2}{2 M} \right\} \;, 
\label{dF-Kolb}
\end{eqnarray} 
where $i=1,2$ corresponds to the field $\Phi,\, S$, respectively. These expressions hold for $\Lambda \ll 1$ and $M^2 \ll m^2 \phi^2$, and only the dominant term has been retained in each entry. 

From Eq.~(\ref{dF-Kolb}) we see that the model approximately meets the conditions (\ref{vs0}) for a vanishing sound speed. The scale of the superpotential, $M$ is much smaller than the inflationary scale, $m$, and $dW/d\phi = 0$. In particular, the inflaton provides a negligible contribution to supersymmetry breaking. 
Applying Eq.~(\ref{vs2}), we find 
\begin{equation}
c_s^2 = \left( \frac{\frac{\dot{\phi}^2}{2} -\frac{m^2 \phi^2}{2} - M^2}{\frac{\dot{\phi}^2}{2} +\frac{m^2 \phi^2}{2} + M^2} \right)^2 + \mathcal{O} \left( \Lambda^2 \right) \;, 
\end{equation} 
which indeed vanishes to $\mathcal{O} \left( \Lambda^0 \right)$ when the inflaton kinetic and potential energy are equal to each other (disregarding the subdominant $M^2$ correction). 
Expanding the various terms to higher order in $\Lambda$ we find that, when the $\mathcal{O} \left( \Lambda^0 \right)$ vanishes, the minimum sound speed is
\begin{equation}
c_{s,{\rm min}}^2 = \frac{256 \, \Lambda^4}{3 \phi^2} \, \left( \frac{M}{m \phi} \right)^{10} \left[ 1 + \mathcal{O} \left( \frac{M^2}{m^2 \phi^2} \right) \right] + \mathcal{O} \left( \Lambda^6 \right) \;. 
\label{vsmin-Kolb}
\end{equation}

The strong stabilization condition $\Lambda \ll 1$ provided us with an expansion parameter to organize the study of the model and present a simple analytical result. However, we see that it (or the nilpotency condition) is not origin for the near vanishing of $c_s$, which is instead mostly due to the absence of $\phi$ from the superpotential and from the smallness of the mass scale $M$. 

To emphasize this, let us consider a different model with strong stabilization, which does not lead to a small gravitino sound speed. The model  \cite{Dudas:2016eej} is  also characterized characterized by the two superfields $\Phi$ and $S$, and by 
\begin{eqnarray} 
K &=&  - \frac{\left( \Phi - {\bar \Phi} \right)^2}{2} + \left\vert S \right\vert^2
- \frac{\left\vert S \right\vert^4}{\Lambda^2} \;, \nonumber\\ 
W &=& f \left( \Phi \right) \left( 1 + \delta \, S \right) 
\;\;,\;\; f \left( \Phi \right) = f_0 + \frac{m}{2} \, \Phi^2  \;. 
\label{mod-Emilian}
\end{eqnarray} 
The K\"ahler potential in this model is relatively simple, and the origin of the inflationary potential resides in the superpotential.
Supersymmetry breaking in the vacuum is due to the constant $f_0$, though supersymmetry breaking gets a contribution from the inflaton during inflation and in its subsequent oscillations.
We can anticipate that the condition (\ref{vs0}), and in particular, 
$\dot{\varphi}\cdot F^* = 0$ will not be satisfied. 

As in the previous example considered, this model has real solutions. The field $s$ is stabilized to a $\phi-$ dependent $\mathcal{O} \left( \Lambda^2 \right)$ value, while the parameter $\delta$ can be set to $\delta = \sqrt{3} + \frac{\Lambda^2}{2 \sqrt{3}} + \mathcal{O} \left( \Lambda^4 \right)$ to ensure that the vacuum energy vanishes in the minimum at $\phi =0$. We then find that the potential and the gravitino mass are given by 
\begin{eqnarray} 
V \left( \phi \right) &=& V^{(0)} \left( \phi \right) + \mathcal{O} \left( \Lambda^2 \right) \;, \nonumber \\
m_{3/2} \left( \phi \right) & = & m_{3/2}^{(0)} \left( \phi \right)  + \mathcal{O} \left( \Lambda^2 \right) \;,
\end{eqnarray}
with
\begin{equation} 
V^{(0)} \left( \phi \right) = 
\left\vert f' \left( \phi \right) \right\vert^2 = \frac{m^2 \phi^2}{2} \;, 
\end{equation}
and
\begin{equation} 
m_{3/2}^{(0)} \left( \phi \right) = \left\vert f \left( \phi \right) \right\vert  = 
f_0 + \frac{m \, \phi^2}{4}  \;,  
\end{equation} 
and we see that $m$ is the inflaton mass, while $f_0$ is the gravitino mass in the vacuum. In terms of the zeroth order potential and gravitino mass, the $\phi-$dependent minimum value of $s$ is given by the compact expression 
\begin{equation}
\left\langle s \right\rangle_\phi = \frac{2 m_{3/2}^{(0)2} - V^{(0)}}{m_{3/2}^{(0)2}} \, \frac{\Lambda^2}{2 \, \sqrt{6}} + \mathcal{O} \left( \Lambda^4 \right) \,. 
\end{equation} 

For this model, we then have 
\begin{eqnarray} 
\dot{\varphi}^i &=& \left\{ \frac{\dot{\phi}}{\sqrt{2}} ,\, 0 \right\} + \mathcal{O} \left( \Lambda^2 \right)  \;, \nonumber\\ 
F^i &=& \left\{ \frac{m \phi}{\sqrt{2}} ,\, 
\sqrt{3} \left( f_0 + \frac{m \, \phi^2}{4} \right) \right\}  + \mathcal{O} \left( \Lambda^2 \right) \;. 
\label{dF-Emilian}
\end{eqnarray} 
We see that $F^\phi$ is not suppressed, and hence we do not expect a small sound speed. Indeed, we obtain 
\begin{eqnarray}
c_s^2 &=& 1 - \frac{6 \, \dot{\phi}^2 \, m_{3/2}^{(0)2}}{\left[ 
\frac{1}{2} \dot{\phi}^2 + V^{(0)} + 3 \, m_{3/2}^{(0)2}   \right]^2} + \mathcal{O} \left( \Lambda^2 \right) \nonumber\\ 
&=& 
\frac{\left[\phi^2 \left( 1 + \frac{3}{8} \, \phi^2 \right) +  \frac{\dot{\phi}^2}{m^2} \right]^2 - \frac{3}{2} \frac{\dot{\phi}^2}{m^2} \phi^4}{\left[ \phi^2 \left( 1 + \frac{3}{8} \, \phi^2 \right) +  \frac{\dot{\phi}^2}{m^2} \right]^2} \nonumber\\ 
& + & \mathcal{O} \left( \frac{f_0}{m} ,\, \Lambda^2 \right) \;,
\end{eqnarray} 
which is always of $\mathcal{O} \left( 1 \right)$. 

The comparison of the two models (\ref{kolb-Lambda}) and (\ref{mod-Emilian}) shows that, as already remarked, the main cause for the suppressed sound speed is the smallness of the inflaton $F-$term. In the model (\ref{kolb-Lambda}) this happens because of the rather peculiar fact that $\Phi$ is absent from the superpotential, particularly when combined with the lack of mobility of $S$ ($S=0$ in the nilpotent limit $\Lambda \to 0$). That is, $\dot{\varphi}$ and $F^*$ are nearly orthogonal ($\dot \varphi^i\sim\{\dot \phi,0 \}$, $F^i \sim \{0,F_S \}$). In the model of (\ref{mod-Emilian}), that is clearly not the case, as there is a large contribution to supersymmetry breaking from $\Phi$ as long
as $\Phi$ is displaced from its minimum at $\Phi = 0$. 
Thus, with this latter simple example, we see that 
$c_s \to 0$ is not a general consequence for supergravity models of inflation, nor for the restricted set of examples with a nilpotent field. 

The above two examples, lead to chaotic inflation models of the type $V = m^2 \phi^2$ \cite{Linde:1983gd}. As this model is disfavored due to the upper limit on the tensor-to-scalar ratio (the model predicts $r = 0.13$, whereas the experimental upper limit is $r < 0.06$ \cite{planck18,rlimit}), we next consider an alternative set of models leading to Starobinsky-like inflation \cite{Staro}.  These models are based on no-scale supergravity \cite{no-scale,LN} which can be derived from string models as their effective low-energy theories \cite{Witten}. 
Further, there is strong relation between no-scale supergravity and higher derivative theories of gravity \cite{Cecotti,DLT,eno9}. 
Starobinsky inflation models in no-scale supergravity generally require two chiral superfields, one of which is associated with the volume modulus of string theory. Depending on a choice of basis, this or the second field may serve as the inflaton \cite{eno7,enov1,building}. In the former, it is necessary to include a third chiral multiplet for supersymmetry breaking.  In the remainder of this section, we consider an example of each type.

In the first Starobinsky-like model we consider, the volume modulus serves as the inflaton \cite{FeKR,eno7}. As noted above, to include supersymmetry breaking, we need to add a third chiral superfield
\cite{egno4,Dudas:2017kfz,Ellis:2020xmk} which we denote as $Z$ and plays the role of a Polonyi field \cite{pol}.\footnote{The model described here is the  untwisted Polonyi field model of ref. \cite{egno4,Dudas:2017kfz,Ellis:2020xmk}. Analogous result are obtained for the twisted Polonyi field model of that work. To keep a notation as close as possible to that of the models (\ref{kolb-Lambda}) and (\ref{mod-Emilian}) discussed above, we have relabeled as $\Phi$ and as $S$ the inflaton and the stabilizer field that were denoted, respectively, as $T$ and $\phi$ in those papers.} The K\"ahler potential and superpotential can be written as
\begin{eqnarray} 
K &=& - 3 \, \ln \left( \Phi + {\bar \Phi} - \frac{1}{3} \left\vert S \right\vert^2 + \frac{\left\vert S  \right\vert^4}{\Lambda_S^2} - \frac{1}{3} \left\vert Z \right\vert^2 + \frac{\left\vert Z  \right\vert^4}{\Lambda^2} \right) \;, \nonumber\\ 
W &=& m \left[ \sqrt{3} \, S \left( \Phi - \frac{1}{2} \right) + \delta \left( Z + b \right) \right] \; . \label{staro1}
\end{eqnarray} 
As we will see, Starobinsky inflation is obtained for $S$ and $Z$ near 0. 

As in the previous cases, the model admits real solutions, and we parametrize as $\Phi = {\bar \Phi} = \phi ,\, S = {\bar S} = s ,\, Z = {\bar Z} = z $. The potential is minimized by $\left\langle \phi \right\rangle = \frac{1}{2} + \frac{\delta^2}{3} + \mathcal{O} \left( \delta^4 ,\, \Lambda^2 \right)$, by $\left\langle s \right\rangle = \delta + \mathcal{O} \left( \delta^3 ,\, \Lambda^2 \right)$, and by $\left\langle z \right\rangle = \frac{\Lambda^2}{6 \sqrt{3}} + \mathcal{O} \left( \delta^2 \Lambda^2 ,\, \Lambda^4 \right)$, where the coefficient $b$ has been set to $b = \frac{1}{\sqrt{3}} \left( 1 - \frac{\delta^2}{6} \right) + \mathcal{O} \left( \delta^4 ,\, \Lambda^2 \right)$ so to have a vanishing potential in the minimum. In the minimum the gravitino mass is 
\begin{equation}
m_{3/2} = \frac{m \, \delta}{\sqrt{3}} + m \times \mathcal{O} \left( \delta^3 ,\, \Lambda^2 \right) \;. 
\end{equation}  
The extremization of the potential with respect to $s$ and $z$ when $\phi$ is away from the minimum gives instead 
\begin{eqnarray} 
\left\langle s \right\rangle_\phi &=& \frac{4 \Lambda_S^2 \phi}{36 \left( 1 - 2 \phi \right)^2 \phi + \Lambda_S^2 \left( 1 + 4 \phi - 4 \phi^2 \right)} \, \delta + \mathcal{O} \left( \delta^3 ,\, \Lambda^2 \right) \;, \nonumber\\ 
\left\langle z \right\rangle_\phi &=& \frac{2 \Lambda^2 \Lambda_S^2 \phi}{3 \sqrt{3} \left[ 36 \left( 1 - 2 \phi \right)^2 \phi + \Lambda_S^2 \left( 1 + 4 \phi - 4 \phi^2 \right) \right]}  \nonumber\\ 
&&\quad\quad + \mathcal{O} \left( \delta^2 \Lambda^2 ,\, \Lambda^4 \right) \;. 
\end{eqnarray} 
The potential along this direction is 
\begin{equation}
V = \frac{3m^2}{16} \, \frac{\left( 1 - 2 \phi \right)^2}{\phi^2} 
+ m^2 \times \mathcal{O} \left( \delta^2 ,\, \Lambda^2 \right) \;. 
\end{equation} 
Note that unlike the previous examples discussed, $\phi$ 
does not have a properly normalized kinetic term.
For small $s$ and $z$, the canonical field is given by
\beq
\label{canrho}
\rho = \sqrt{\frac32} \ln \left( 2 \phi \right) \, ,
\eeq
so that the potential becomes
\beq
V = \frac{3 m^2}{4} \left(1-e^{-\sqrt{2/3}\rho} \right)^2 \, .
\label{staropot}
\eeq
The inflaton mass is $m$ and hence we see that $\delta \ll 1$ is the ratio of the gravitino mass to the inflaton mass and justifies the  limit that we have adopted in our analysis. 
Along this direction, labeling as $1,2,3$ the fields $\Phi ,\, S ,\, Z$, respectively, we find 
\begin{eqnarray}
\dot{\varphi}^i &=& \left\{ \dot{\phi} ,\, m \times \mathcal{O} \left( \delta ,\, \Lambda^2 \right) ,\, m \times \mathcal{O} \left( \Lambda^2 \right) \right\} \;, \nonumber\\ 
F^i &=& m \Bigg\{ \mathcal{O} \left( \delta ,\, \Lambda^2 \right) ,\,- \frac{\sqrt{3} \left( 1 - 2 \phi \right)}{\sqrt{2 \phi}} + \mathcal{O} \left( \delta^2 ,\, \Lambda^2 \right) ,\, \nonumber\\ 
&& \quad\quad\quad\quad\quad\quad\quad\quad\quad\quad
\mathcal{O} \left( \delta ,\, \Lambda^2 \right) \Bigg\} \;. 
\end{eqnarray} 
We see that $\dot{\varphi} \cdot F$ is suppressed, leading to the possibility of a small gravitino sound speed. We indeed find 
\begin{equation}
c_s^2 = \left( \frac{4 \dot{\phi}^2-m^2\left(1-2 \, \phi \right)^2}{4 \dot{\phi}^2+m^2\left(1-2 \, \phi \right)^2} \right)^2 + \mathcal{O} \left( \delta^2 ,\, \Lambda^2 \right) \;, 
\end{equation} 
and the leading term can vanish during the oscillations of the inflaton. 

We conclude this section with a second no-scale supergravity model, based on a simple Wess-Zumino form for the inflaton superpotential \cite{eno6}. In this case, the volume modulus is strongly stabilized \cite{eno7}. This simple model further emphasizes that the suppression of the sound speed is by no means a generic feature of supergravity models. The model does not require an additional field for supersymmetry breaking and can be written as \cite{egno4,Ellis:2020xmk}
\begin{eqnarray}
K \!\!&=&\!\! - 3 \, \ln \left( S + {\bar S} + \frac{\left( S + {\bar S} - 1 \right)^4}{\Lambda^2} + \frac{d \left( S - {\bar S} \right)^4}{\Lambda^2} - \frac{\left\vert \Phi \right\vert^2}{3} \right) \;, \nonumber\\ 
W \!\!&=&\!\! m \, \left(\frac{\Phi^2}{2} - \frac{\Phi^3}{3 \sqrt{3}}\right) + \lambda \, m^3 \;, 
\end{eqnarray} 
where $d$ is a constant of $\mathcal{O}(1)$, and $\lambda$ sets the scale of supersymmetry breaking.  

The model admits real solutions that we parametrize as in the previous models, $\Phi = {\bar \Phi} = \phi ,\, S = {\bar S} = s$. In the minimum, $\left\langle \phi \right\rangle = 0$ and $\left\langle s \right\rangle = \frac{1}{2}$. For these values the potential vanishes, while the gravitino mass is $m_{3/2} = \lambda \, m^3$. 

Extremizing the potential with respect to $s$, with the inflaton away from the minimum, we find 
\begin{equation}
\left\langle s \right\rangle_\phi = \frac{1}{2} + 
\frac{\phi^2}{2 \left( 6 \lambda m^2 + \phi^2 \right)^2} \, \frac{\sqrt{3}-\phi}{\sqrt{3}+\phi} \Lambda^2 + \mathcal{O} \left( \Lambda^4 \right) \;,  
\end{equation} 
leading to the potential 
\begin{equation}
V = \frac{3 m^2 \phi^2 }{\left( \sqrt{3}+\phi\right)^2}  + \mathcal{O} \left( \Lambda^2 \right) \;. 
\end{equation} 
As in the previous no-scale example,
$\phi$ is not canonical and writing
\beq
\phi = \sqrt{3} \tanh \left( \frac{\rho}{\sqrt{6}} \right)
\eeq
leads to the same potential given in Eq.~(\ref{staropot}).

We then find, using the same notation as for the previous models,  
\begin{eqnarray}
\dot{\varphi}^i &=& \left\{ \dot{\phi} ,\, m \times \mathcal{O} \left( \Lambda^2 \right) \right\} \;, \nonumber\\ 
F^i &=& \left\{ 
m \phi \, \sqrt{\frac{\sqrt{3}-\phi}{\sqrt{3}+\phi}} ,\, 
- \frac{m \left( 3 \lambda m^2 + \phi^2 \right)}{2 \sqrt{3} \sqrt{3-\phi^2}} \right\} \nonumber \\
&& \quad\quad\quad\quad\quad\quad\quad\quad\quad\quad + m \times \mathcal{O} \left( \Lambda^2 \right) \; .
\end{eqnarray} 
We see that $\dot{\varphi} \cdot F$ is not suppressed, and we do not expect a suppression in the gravitino sound speed. A direct inspection of the $\mathcal{O} \left( \Lambda^0 \right)$ term indicates that this is indeed the case, though the full expression is not particularly illuminating. In the $\sqrt{\lambda} m \ll \phi ,\, \frac{\dot{\phi}}{m} \ll 1$ regime we obtain 
\begin{equation}
c_s^2 \simeq 1 - \frac{m^2 \phi^4 \dot{\phi}^2}{6 \left( m^2 \phi^2 + \dot{\phi}^2 \right)^2} \;,\;\; \Lambda \ll 1 \;,\; \sqrt{\lambda} m \ll \phi ,\, \frac{\dot{\phi}}{m} \ll 1 \;. 
\end{equation}

\section{Complete linearized equation for the longitudinal gravitino}
\label{sec:theta-Y}

So far, we have seen that while it is possible to find models with $c_s \simeq 0$, it is by no means a generic feature of supergravity inflation models. Next, we point out that even
if $c_s = 0$, one can not conclude that there is any enhanced production of gravitinos. 
We show that this is the case by considering the equation of motion for the longitudinal gravitino, as obtained in \cite{Kallosh:2000ve} and then studied in \cite{Nilles:2001fg}. 

In the unitary gauge, the dynamical variable encoding the longitudinal gravitino is the combination $\theta = \gamma^i \psi_i$, where $\psi$ is the gravitino field. In a nontrivial background, the longitudinal gravitino is coupled to another fermionic combination $\Upsilon$ at the linearized level. In a cosmological background, where the scalar fields depend on time, 
\begin{equation}
\Upsilon = K_{ij^*} \left( \chi^i \partial_0 \varphi^{j^*} +  \chi^{j^*} \partial_0 \varphi^i \right) \;, 
\end{equation} 
where the indices run over the number of chiral superfields in the model, $\left( \varphi^i ,\, \chi^i \right)$ are the scalar and fermion components of a chiral complex multiplet (where 
$\chi^i$ is a left-handed fermion), while $\left( \varphi^{i^*} ,\, \chi^{i^*} \right)$ are their conjugates. 

For simplicity, we consider the case of two chiral superfields, although this can be immediately generalized to an arbitrary number. In this case, the coupled equations for $\theta$ and $\Upsilon$ form a closed system. To write these equations, we introduce the following quantities (that we express through the scalar product defined above)
\begin{eqnarray} 
\alpha &\equiv& \rho + 3 m_{3/2}^2 = \dot{\varphi} \cdot \dot{\varphi}^* + F \cdot F^* \;, \nonumber\\ 
\alpha_1 &\equiv& p - 3 m_{3/2}^2 = \dot{\varphi} \cdot \dot{\varphi}^* - F \cdot F^* \;, \nonumber\\ 
\alpha_2 &\equiv& 2 \frac{\partial}{\partial t} 
\left[ {\rm e}^{K/2} \, W \, P_L + {\rm e}^{K/2} \, W^* \, P_R \right] \nonumber\\ 
&=& 2 \left( \dot{\varphi} \cdot F^* \, P_L + 
\dot{\varphi}^* \cdot F \, P_R \right) \nonumber\\ 
&& \quad\quad + {\rm e}^{K/2} \left( \dot{\varphi}^i \frac{\partial K}{\partial \varphi^i} -  \dot{\varphi}^{*i} \frac{\partial K}{\partial \varphi^{*i}} \right) \left( W^* \, P_R - W \, P_L \right) \;. \nonumber\\ 
\end{eqnarray} 
In these expressions $\rho$ and $p$ denote, respectively, the energy density and pressure of the background scalars, while $P_{L/R}$ are the left- and right-handed chiral fermion projection operators. These expressions considerably simplify along real solutions for the scalar fields 
\begin{eqnarray} 
{\rm real \; scalars}: \;&& \alpha = 
\dot{\varphi} \cdot \dot{\varphi} + F \cdot F \;, \nonumber\\ 
&& \alpha_1  = \dot{\varphi} \cdot \dot{\varphi} - F \cdot F \;, \nonumber\\ 
&& \alpha_2 = 2 \, \dot{\varphi} \cdot F \;. 
\label{alphas-real}
\end{eqnarray} 
These expressions apply to the models considered in the previous sections, that have real background solutions. In the following we employ the simplified expressions (\ref{alphas-real}). 

The closed system of equations for $\theta$ and $\Upsilon$ is 
\begin{equation} 
\left( \gamma^0 \, \partial_0 + i \gamma^i \, k_i \, N + M \right) X = 0 \;\;\;\;\;,\;\;\;\;\; X = \left( \begin{array}{c} 
{\tilde \theta} \\ {\tilde \Upsilon} 
\end{array} \right) \;, 
\label{eqs-theta-Y} 
\end{equation} 
where ${\tilde \theta}$ and ${\tilde \Upsilon}$, are canonically-normalized fields, related to the original fields by 
\begin{equation}
\theta \equiv \frac{2 i \gamma^i \, k_i}{\left( \alpha \, a^3 \right)^{1/2}} \, {\tilde \theta} \;\;\;,\;\;\; 
\Upsilon \equiv \frac{\Delta}{2} \left( \frac{\alpha}{a} \right)^{1/2} \, {\tilde \Upsilon } \;,  
\end{equation} 
(with $\Delta$ to be defined shortly), and where 
\begin{equation} 
N = \left( \begin{array}{cc} 
- \frac{\alpha_1}{\alpha} - \gamma^0 \, \frac{\alpha_2}{\alpha} & - \gamma^0 \, \Delta \\ 
- \gamma^0 \, \Delta & - \frac{\alpha_1}{\alpha} + \gamma^0 \, \frac{\alpha_2}{\alpha} 
\end{array} \right) \;,  
\label{N}
\end{equation} 
while the expression for $M$ is not important for the present discussion, and can be found in \cite{Nilles:2001fg}. 

Disregarding the presence of the field $\Upsilon$ amounts to the system studied in 
\cite{Kolb:2021xfn,Kolb:2021nob}. The square of the gravitino sound speed would then be given by the square of the $N_{11}$ element (we note that, due to the signature we have chosen, $\gamma^0$ is anti-hermitian), 
\begin{equation}
N_{11} \, N_{11}^\dagger = \frac{\left\vert \alpha_1 \right\vert^2 + \left\vert \alpha_2 \right\vert^2}{\alpha^2} = c_s^2 \;. 
\end{equation} 
Namely, using the expressions (\ref{alphas-real}) leads precisely to the sound speed (\ref{vs2}). The complete system however has also off-diagonal elements, with %
\begin{equation}
\Delta = \sqrt{1-c_s^2} \;. 
\end{equation} 
Therefore, when the coefficient $c_s^2$ vanishes, 
\begin{equation}
c_s^2 = 0 \;\; \to \;\; \Delta = 1 \;\;,\;\; N =  \left( \begin{array}{cc} 
0 & - \gamma^0  \\ 
- \gamma^0  & 0 
\end{array} \right) \;,  
\end{equation} 
leading to a non-singular matrix $N$, and therefore to a nonvanishing sound speed for the physical eigenstates of the system. 
Consequently {\em none} of the models discussed in Section 3 have catastrophic production of gravitinos. Problems can only arise if $\Upsilon$ can be ignored as is the case when a second, orthogonal nilpotency condition, is applied as discussed in the next section. Then, the only problematic models would be those defined in Eqs.~(\ref{kolb-Lambda}) and (\ref{staro1}), in the case where the inflaton multiplet $\Phi$ is subject to the additional orthogonal constraint $S ({\Phi} - \overline{\Phi}) = 0$. Indeed, this additional constraint removes the inflatino from the spectrum (hence $\Upsilon = 0$) and the speed of sound in these models hits zero at some point during the inflationary evolution. As we will see, such models seem suspicious from the viewpoint of a fundamental theory of gravity.

\section{Models with Orthogonal Nilpotent Superfields}
\label{sec:nil}

In the cases considered in this paper, the goldstino $G$ belongs to a chiral multiplet and has a scalar superpartner (the sgoldstino), which, once supersymmetry is broken,  acquires a non-supersymmetric mass. 
Decoupling the sgoldstino by giving it an {\it infinite} mass leads to  a non-linear realization of supersymmetry. 
A particularly simple way non-linear realization can be obtained is by imposing the nilpotent constraint
\cite{Rocek:1978nb,Komargodski:2009rz,Antoniadis:2014oya,Ferrara:2014kva,Kallosh:2014via,DallAgata:2014qsj,Dudas:2015eha,Ferrara:2015gta,Ferrara:2015cwa}
\begin{equation}
\label{X2}
S^2 = 0 \, . 
\end{equation} 

When supersymmetry is broken by means of a non-trivial $F^S \neq 0$, the constraint is solved by
\begin{eqnarray}
S = \frac{G^2}{2F^S} + \sqrt 2 \theta G + \theta^2 F^S \, . 
\label{nil1}
\end{eqnarray}
Here and in what follows, we discuss the constraints at the level of global supersymmetry for simplicity. The generalization to supergravity can be found in the literature \cite{Antoniadis:2014oya}-\cite{DallAgata:2016syy}. 
The constraint (\ref{X2}) can be interpreted as the infrared limit of a very heavy sgoldstino. This can be obtained starting from a microscopic Lagrangian of the type \cite{Komargodski:2009rz}:
\begin{equation}
K = |S|^2 - \frac{1}{\Lambda^2} |S|^4 \, , \qquad \ W = W_0 + W_1 S \, ,  \label{nil2}    
\end{equation}
in the limit $\Lambda \to 0$. Indeed, the sgoldstino mass
$m_S^2 = 4 \frac{{F^S}^2}{\Lambda^2}$ is sent to infinity in this limit, leading to a nonlinear realization of supersymmetry in the IR. The limit $\Lambda \to 0$ has its limitations
\cite{Dudas:2016eej}, since it implies  field-theory dynamics in some heavy sector, which after decoupling, leaves behind the ``strong stabilization" term $\frac{1}{\Lambda^2} |S|^4$. Modulo these subtleties, the UV Lagrangian  (\ref{nil2}) contains only two derivatives and is pretty standard. 

The situation is different for the orthogonal constraint on the chiral superfield ${\Phi}$ that removes the imaginary part of the scalar, the fermion, and the auxiliary field \cite{Komargodski:2009rz,Ferrara:2015tyn,Carrasco:2015iij,DallAgata:2015zxp,Kallosh:2016hcm,DallAgata:2016syy}:
\begin{equation}
S ({\Phi} - \overline{\Phi}) = 0 \ . \label{nil3}
\end{equation}
It was shown in \cite{DallAgata:2016syy}  that (\ref{nil3}) is equivalent to the following set of constraints
\begin{eqnarray}\label{nil4}
 |S|^2 ({\Phi} - \overline{\Phi} )&=& 0 \, ,  \\[2mm]
\label{nil5}
 |S|^2 \overline D_{\dot \alpha} \overline{\Phi} &=& 0\, , \\[2mm]
 \label{nil6}
 |S|^2 \overline D^2 \overline{\Phi} &=& 0 \, .
\end{eqnarray}
Each constraint above eliminates one component field:
Eq.~(\ref{nil4}) eliminates the scalar, Eq.~(\ref{nil5}) eliminates the fermion, whereas Eq.~(\ref{nil6}) eliminates the auxiliary field in the $\Phi$ multiplet. 
The constraint \eqref{nil3} can be obtained starting from a microscopic Lagrangian containing three additional terms \cite{DallAgata:2016syy}, which generate non-supersymmetric masses for the component fields that we remove:
\begin{eqnarray}
&& 	 \int \! d^4 \theta \Big{[} \frac{m^2_b}{2f^2} |S|^2 ({\Phi} - \overline{\Phi} )^2 
	- \frac{g_{F^{\Phi}}}{f^2}  |S|^2 D^2 {\Phi} \overline D^2 \overline {\Phi } \Big{]} \nonumber \\ 
&& 	- \frac{m_\zeta}{2 f^2} \int \! d^4 \theta \Big{[}  |S|^2  D^{\alpha} {\Phi} D_\alpha {\Phi} + c.c. \Big{]} \ . \label{nil7}
\end{eqnarray} 
In (\ref{nil7}), $f$ can be taken to be (by convention) the supersymmetry breaking scale in the vacuum $f= \langle F^S \rangle$, $m_b$ and $m_{\zeta}$ are mass parameters for the decoupling scalar and fermion, respectively, $g_{F^{\Phi}}$
is a dimensionless coupling, $D_\alpha$ denotes a covariant derivative in superspace and $D^2 =D^\alpha D_\alpha$, see e.g. \cite{Wess:1992cp}. 
Notice that only the first term in (\ref{nil7}) is a standard correction to the K\"ahler potential, whereas the second and the third terms contain higher-derivatives. 
The limit $m_b \to \infty$, $g_{F^{\Phi}} \to \infty$ and $m_\zeta \to \infty$ generate infinitely large masses to the scalar and fermion and an infinite coefficient to the auxiliary field $F^{\Phi}$. In this limit,  the superspace equations of motion for the chiral superfield ${\Phi}$ are dominated by 
\begin{eqnarray} 
&& 	\overline D^2 \left\{ \frac{m^2_b}{f^2} |S|^2 ({\Phi} - \overline{\Phi} )
	+\frac{m_\zeta}{f^2} D^\alpha ( |S|^2 D_\alpha {\Phi} ) \right.
	\nonumber \\
&& \quad\quad \left.	- \frac{g_{F^{\Phi}}}{f^2} D^2 (|S|^2 \overline D^2 \overline {\Phi}) \right\}  = 0 \ .\label{nil8}
\end{eqnarray} 
This indeed reproduces the constraints (\ref{nil4})-(\ref{nil6}) and therefore  (\ref{nil3}): multiplication with $S \overline S$ leads to (\ref{nil6}); the multiplication with $D_\beta S \overline S$ gives (\ref{nil5}); and finally, using (\ref{nil6}) and (\ref{nil5}) in (\ref{nil8}) then multiplying with $\overline S$ leads to (\ref{nil4}).
   
As emphasized already,  in contrast to the nilpotent UV Lagrangian 
(\ref{nil2}), the action (\ref{nil7}) contains higher derivatives. As such, it is qualitatively different and the orthogonal constraint cannot therefore be obtained, to our knowledge, starting from a standard two-derivative action in the UV. In particular, if the chiral superfield $S$ was not nilpotent, the second term in Eq. (\ref{nil7}) proportional to $g_{F^\Phi}$ could  introduce ghosts into the theory. This is a special property of the decoupling procedure, and could signal that such constraints could (but not necessarily) come from a sick UV theory. 
This could explain the problems with the slow gravitinos in some inflationary models using the orthogonal constraint. 

We conclude this section with some general properties and implications of supersymmetric models with a second degree nilpotency condition, $S^2 = 0$, and the orthogonality constraint, $S(\Phi - \bar{\Phi}) = 0$. It follows from these constraints that if we choose to work in the unitary gauge the goldstino and inflatino fields are absent in these models. If we consider a supergravity model with a K\"ahler potential and superpotential of the form
\begin{equation}
    K = -\frac{\left(\Phi - \bar{\Phi} \right)^2}{2} + S \bar{S} \, ,
\end{equation}
and
\begin{equation}
    \label{sup1}
    W (\Phi, S) \; = \; f(\Phi) S + g(\Phi) \, , 
\end{equation}
so that the K\"ahler potential possesses a shift symmetry, and vanishes when the field constraints are imposed, $K(S, \bar{S}; \Phi, \bar{\Phi})|_{S = \bar{S} = \Phi - \bar{\Phi} = 0} = 0$. The orthogonality constraint ensures that the covariant derivative $D_{\Phi}W$ is absent in the effective scalar potential, which is then given by \cite{Ferrara:2015tyn,Carrasco:2015iij}
\begin{equation}
    \label{pot1}
    V  \; = \; e^{K} \left( K^{S S^{*}} |D_S W|^2 - 3|W|^2  \right) = f^2 - 3 m_{3/2}^2 \, ,
\end{equation}
where the gravitino mass is 
\begin{equation}
    \label{gravmass}
    m_{3/2} \; = \; |g(\Phi)| \, .
\end{equation}

Note that the sound speed given in Eq.~(\ref{vs2}) should not be used in this case, since $F^\phi = 0$ for the bosonic part, while $D_\phi W \ne 0$.  Rather, one can use
Eq.~(\ref{Eq:generic32}) and the potential in Eq.~(\ref{pot1}) with Eq.~(\ref{gravmass}) to obtain
\begin{eqnarray}
    c_s^2 & = & \frac{\left(f(\Phi)^2 -\dot{\Phi}^2 \right)^2}{\left(f(\Phi)^2 + \dot{\Phi}^2 \right)^2} + \frac{(2 g'(\Phi) \dot{\Phi})^2}{\left(f(\Phi)^2 +\dot{\Phi}^2 \right)^2} \nonumber \\
    & = & 1- 4 \frac{\dot{\Phi}^2}{(f^2 + \dot{\Phi}^2)^2} \left( f^2 - g'^2 \right)
    \, ,
      \label{soundspeed}
\end{eqnarray}
where $g'(\Phi) = \partial g/ \partial \Phi$. If we want to ensure that the sound speed $c_s > 0$ does not vanish, we must impose the constraint
\begin{equation}
    \label{conspos}
    \left( f(\Phi)^2 - \Dot{\Phi}^2 \right)^2 + \left(2 g'(\Phi) \dot{\Phi} \right)^2 > 0 \, .
\end{equation}
The above quantity is non-negative  at all times in moduli space and this condition is only violated when $f(\Phi) = \pm \dot{\Phi}$ and $g'(\Phi) = 0$. If the gravitino mass is constant at all times, this implies that $g'(\Phi) = 0$ and the constraint~(\ref{conspos}) will be violated when the first term vanishes.

While it is straightforward to construct models for which the gravitino mass is not constant,
additional problems quickly ensue. Consider, for example, a superpotential defined by
\begin{equation}
    \label{funcs1}
    f = f_0 + \sqrt{3}g, \qquad g = \frac{1}{2\sqrt{3}} \left(a f_0  + \tilde{m} \right) \, ,
\end{equation}
where $\tilde{m}$ is associated with the weak scale, and the parameter $a$ is an arbitrary positive number. At the minimum of the potential, the gravitino mass is of order ${\tilde m}$. Any variation in a quantity this small will still lead to a sound speed which is also very small. For $a \ne 0$, the variation of the gravitino mass may be comparable to the potential.  The effective scalar potential~(\ref{pot1}) becomes
\begin{equation}
    \label{pot2}
    V = (a + 1) f_0^2 + \tilde{m} f_0 \, .
\end{equation}
If we choose
\begin{equation}
 f_0  =  \frac{\sqrt{3}}{2 \sqrt{a + 1}} m \left(1 - e^{-\frac{2}{\sqrt{3}}\Phi} \right) \, ,
\end{equation}
and use the canonically-normalized field $\Phi = \phi/\sqrt{2}$, we recover approximately, for $\tilde m \ll m$, the Starobinsky inflationary potential~(\ref{staropot}).
However, during each oscillation, when $\Phi = 0$,
$f^2 - g'^2 = -a^2/12(1+a)$.
As a result the sound speed exceeds 1, signalling a sickness in the theory. 
The Cauchy-Schwarz inequality noted above is no longer
guaranteed to hold as the sound speed is now
\begin{equation}
c_s^2 = 1 - \frac{4}{
\left(\vert\dot{\Phi}\vert^2 + \vert F^S \vert^2 \right)^2} \, 
\left\{ 
 |\dot{\Phi}|^2  |F^S|^2  - 
\left\vert \dot{\Phi}  e^{K/2} D_{\Phi} W  \right\vert^2 \right\}  \;, 
\label{vs3}
\end{equation} 
and there is no such inequality between $|\dot{\varphi}|^2  |F|^2 = |\dot{\Phi}|^2 |F^S|^2$ and $\left\vert \dot{\varphi}  \cdot F^*  \right\vert^2 = |e^{K/2} \dot{\Phi} D_\Phi W|^2 $, 
 since we can no longer use Eq.~(\ref{DWF}) to relate $F^\Phi$ and $D_\Phi W$.
Thus pathologies might arise whenever $|e^{K/2} D_\Phi W| > |F^S|$.
Such behavior was also seen in \cite{Kolb:2021xfn}.  We are not sure if this behavior is a remnant of the lack of two derivative UV origin stemming from the orthogonal constraint.

\section{Conclusions}
\label{sec:conclusions}

If supersymmetry is realized below the Planck scale, it is natural to construct models of inflation in the context of supersymmetry/supergravity. Particle production and reheating after the period of exponential expansion is an essential component of any model of inflation, and in the context of 
supersymmetry, the production of gravitinos must be considered.
If the gravitino is not the lightest supersymmetric particle, it is expected to be unstable. Because of its gravitational coupling, its lifetime can be quite long (for weak scale gravitino masses), causing a host of potential problems \cite{sw}.

In general, the production of gravitinos following inflation is complicated by the possibility that the field content of the longitudinal mode comprised of the goldstino may be different during inflation from that in the vacuum. Indeed,
supersymmetry might be broken during inflation, 
and the longitudinal component of the gravitino may be identified with the inflatino. Thus, models
with substantial non-perturbative particle production \cite{Kallosh:1999jj,Giudice:1999yt,Giudice:1999am,Kallosh:2000ve}, may be producing an inflatino rather than a gravitino \cite{Nilles:2001ry,Nilles:2001fg}.
The inflatino problem is generally model dependent
and is non-existent in certain constructions \cite{Nilles:2001my}.

An additional problem associated with gravitinos has been discussed recently \cite{hase,Kolb:2021xfn,Kolb:2021nob} in connection with a vanishing sound speed for gravitinos.
It was argued that when $c_s \to 0$, the non-perturbative production
of high-momentum modes in the helicity 1/2 mode is unsuppressed.  Indeed, there are a variety of well-studied models where the gravitino sound speed does vanish. We have discussed several examples for which $c_s \to 0$ and several for which it does not. That is, while a vanishing sound speed may occur in supergravity models of inflation, it is by no means a necessary consequence. Furthermore, we have shown that even in the models where $c_s \to 0$, runaway gravitino production is blocked by mixing with the inflatino in a time-dependent background. This is the case in models with and without a nilpotent field associated with supersymmetry breaking.  

However, the problem of catastrophic gravitino production may occur in models with a second orthogonal constraint for which the inflatino is removed from the particle spectrum. It is not clear how these models can be extracted from a UV theory involving no more than two derivatives. This argument along with the problem of a vanishing sound speed (or superluminal sound speed when $c_s \ne 0$) may be a constraint on models with constrained fields rather than supergravity models of inflation in general.

\vskip.1in
{\bf Acknowledgments:}
\noindent 
The authors want to thank T. Gherghetta, H.P. Nilles, and L. Sorbo for helpful discussions. This project has received support from the European Union’s Horizon 2020 research and innovation programme under the Marie Sk$\lslash$odowska-Curie grant agreement No 860881-HIDDeN, by the ANR grant Black-dS-String ANR-16-CE31-0004-01, the CNRS PICS MicroDark, and the IN2P3 Master Project UCMN.
The work of K.A.O.~was supported in part by DOE grant DE-SC0011842  at the University of
Minnesota. The work of MG was supported by the Spanish Agencia Estatal de Investigaci\'on through Grants No.~FPA2015-65929-P (MINECO/FEDER, UE) and No.~PGC2018095161-B-I00, IFT Centro de Excelencia Severo Ochoa SEV-2016-0597, and Red Consolider MultiDark FPA2017-90566-REDC.

\end{document}